\documentstyle[prl,aps,psfig]{revtex}
\begin{document}
\draft

\twocolumn[\hsize\textwidth\columnwidth\hsize\csname
@twocolumnfalse\endcsname

\widetext
\title{ From antiferromagnetism to d-wave superconductivity in  
    the $2 D$ $t-J$ model}
\author {Matteo Calandra and Sandro Sorella }
\address{ Istituto Nazionale di Fisica della Materia and
  International School for Advanced Studies
Via Beirut 4, 34013 Trieste, Italy }
\date{\today}
\maketitle
\begin{abstract}
We have found that the two dimensional t-J model, for the 
physical parameter range $J/t = 0.4$   reproduces 
the main experimental qualitative features of High-Tc copper oxide
superconductors: d-wave superconducting correlations are strongly 
enhanced upon small doping and clear evidence of off diagonal long 
range order is found at the optimal doping $\delta \approx 0.15$. 
On the other hand antiferromagnetic  long range order, 
clearly present at zero hole doping, is  suppressed
 at small hole density  with clear absence of   
antiferromagnetism   at $\delta >  \sim 0.1$. 
\end{abstract}
]

\narrowtext

The interplay between antiferromagnetism and
superconductivity in the $CuO_2$ layers of the high $T_c$
compounds is one of the most important effects  where strong
electron correlation may play the main
role.\cite{zhang,Zhang88} However, after many years of
theoretical studies and experimental efforts \cite{Dagottone} the most
obvious question  is still unclear:
 whether the occurrence of high Tc superconductivity  is determined by the
proximity of the compound to a perfect antiferromagnetic insulator.

In case strong correlation is the dominant force driving  from antiferromagnetism
to superconductivity a well accepted model  is the 2D  t-J model\cite{Zhang88}:
\begin{equation}
\label{tjmodel} H = J \sum_{<i,j>} ( S_i \cdot S_j
- \frac{1}{4} n_i n_j ) - t \sum_{<i,j>} (
c^{\dagger}_{i \sigma} c_{j \sigma} + h.c. ),
\end{equation}
where $c^{\dagger}_{i \sigma}$ creates an electron  of spin
$\sigma$ on the site $i$, $n_i$ and $S_i$ being the electron
number and spin operators respectively. Double occupations  are
forbidden  and $< i,j >$ denotes  nearest neighbor summation
 over the $L$ lattice sites with periodic boundary conditions.

In the last decade the investigation of the properties of the $2
D$ $t-J$ model (and of the parent Hubbard model) has been a challenge  
for numerical calculations. Exact diagonalization (ED)
\cite{Dagotto92} shows that antiferromagnetic correlations are
resistant up to $\delta \sim 0.15$ and superconductivity is present at
intermediate doping but the lattice sizes
considered were too small for being conclusive. On the
contrary the quantum Monte Carlo (QMC) methods allow simulations
on larger systems but suffer from the well known ``minus sign
problem" instability, which makes the simulation 
 impossible at low enough  temperatures.

 At present, this instability  can be controlled,
only at the price of introducing some approximation, such as
 the fixed node (FN) approximation
\cite{Haaf95}, which is strictly variational on the ground
state  energy, the constrained path quantum Monte Carlo\cite{Zhang95}
and the Green function Monte Carlo with stochastic reconfiguration
GFMCSR\cite{Sorella98}, which has been developed  to improve the
accuracy of the FN. Both the FN and GFMCSR techniques will be
extensively used in this work. Similar approximations on the
ground state wavefunction can be obtained by applying one (or
more) Lanczos steps (LS) to the variational wavefunction
\cite{Heeb94,khono,Shih98}, or also using the density matrix 
renormalization group (DMRG), which in 2D is also
affected by a sizable error, and is not ``numerically exact'' as
in 1D.

All these approximations  allow  to obtain a rather
accurate value of the ground state energy of the model,
 with an error typically  less than $1\%$ of the correlation energy even
for  large $L$. However this kind of accuracy
for the energy certainly does not allow to draw reasonable
conclusions on  the interesting long range properties of
the model, see e.g. \cite{Shastry97}. On the other hand it is reasonable to
expect that, by using approximate techniques that do not spoil  the local
character of the Hamiltonian,
 a similar good accuracy can be obtained on the ground state
expectation value of  short range  operators
like, for instance,
 the kinetic energy and the  exchange energies in Eq.\ref{tjmodel}.
These  operators $O$, 
acting only on nearest neighbor sites, share the important
property that, if  added to the Hamiltonian ( $ H_h  \to H- h O$ )  do
not change its  local character. Moreover
 this kind of perturbation typically leads to 
  a sizable change of
the ground state energy per site $E_h$  even in the linear regime
 $E_h = E_0 - h  <O>/L  +o(h)$, providing a very reliable estimate 
of the ground state expectation value $<O>$, as the energy $E(h)$ can 
be accurately determined for few values of the field $h$.

So far, in the literature\cite{Zhang97,White,Shih98},  the ground state expectation value of the squared 
order parameter $O^2$ is  estimated on an approximate ground state $\tilde \psi_0$,
by taking, simply,  its  bare expectation value $<\tilde \psi_0|O^2|\tilde \psi_0>$.
For long range operators such as $O^2$,
 this may lead to  very poor approximations, unless the method is almost
exact.

In order to detect superconducting long range order with a more controlled
 approximation, we perform simulations in the grand canonical ensemble
 and add to $H$ a short range operator  which creates or destroys a $d$-wave singlet 
Cooper pair:  
\begin{equation} \label{eq:Hpert}
H(h)= H - h \left(\Delta^+ + \Delta \right) - \mu \hat N
\end{equation}
where $\Delta^+ =\sum_{<i,j>}M_{ij} (c^+_{i \uparrow}c^+_{j
\downarrow} + c^+_{j \uparrow}c^+_{i \downarrow} ) $
 and $M_{ij}=1$ or $-1$ if the bond $<i,j>$ is in the $x$ or $y$
direction respectively,  
while $\mu$ is the chemical potential and $\hat N$ the particle operator. 
FN and GFMCSR allow to compute quite accurately the ground state energy $E(h)$
also in presence of 
the field $h$. To this purpose a fundamental role is played by the guiding 
wavefunction which allows to perform importance sampling.
 We generalize the $N$
particle, d-wave symmetry, BCS guiding wavefunction\cite{Gros88}
($|BCS\rangle$)  to the grand
canonical ensemble  by introducing a proper weight $f_N$ for each $N$ 
particle sector:
\begin{equation}\label{eq:psig}
|\psi_G \rangle = \sum_N f_N P_N {\cal P}_G 
|BCS\rangle
\end{equation}
where ${\cal P}_G$
projects out  doubly occupied sites and $P_N$ selects the  
  $N-$particle component of the wavefunction.
\begin{figure}
\centerline{\psfig{figure=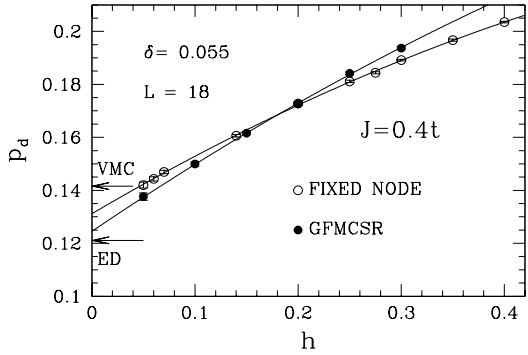,width=7cm}}
\caption{\baselineskip .185in  $p_d$ for $N=16$,
$L=18$}\label{fig:par18}
\end{figure}
Our purpose is to  compute the anomalous average of the order parameter
$p_d= |\langle N+2|\Delta^+|N\rangle|/L$,
where  $|N\rangle$ and $|N+2\rangle$
are the $N$ and $N+2$ particle ground state respectively.
$p_d$ can be non zero even on a finite size 
and zero external field.
Moreover, if superconducting long range order occurs,
$p_d$  remains  finite for $L \to \infty$.

In order to compute $p_d$ on  finite size  systems
we have implemented the following  simple strategy.
We choose the chemical potential $\mu$ in
 a way that the ground state energies per site $E_N$ and $E_{N+2}$ for the $N$ and $N+2$ 
particles are degenerate. 
In order to reduce the ground state energy statistical error 
 we optimize the variational parameters $f_N$ 
by  restricting ourselves to the subspaces  of $N$ and $N+2$ particles 
relevant for the matrix element $p_d$, $f_N$ being zero otherwise. 
In the guiding function $f_N$ and $f_{N+2}$  are then  determined 
by requiring that  the average particle number  $<\psi_G| \hat N|\psi_G>$
is  equal to $N+1$.
 The first order correction to the energy due to the perturbation  
(\ref{eq:Hpert}) in this restricted Fock space is given by the eigenvalues 
of the secular matrix:
\begin{equation}
\begin{array}{|cc|}
  E_N  & \pm h p_d \\
  \pm h p_d & E_{N+2} .
\end{array}
\end{equation}
 It easily follows that  $ E(h)=E_N - h p_d$, meaning
that the anomalous average of the order parameter can be computed
as an energy difference $(E_N - E(h))/h$
for $h\to 0$.
 {\em A long range property of the model
 can be probed by studying
 the ground state energy change under the effect of a
local perturbation.}
We expect this scheme to be  a much more controlled and accurate  way 
to characterize the long distance behavior   of a model. 
\begin{figure}
\centerline{\psfig{figure=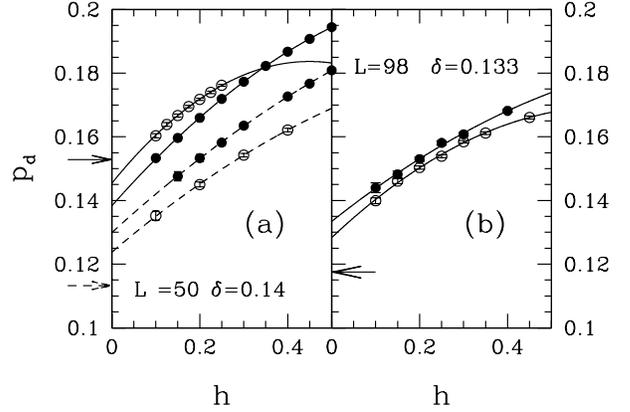,width=8cm}}
 \caption
{\baselineskip .185in VMC(arrows), FN(empty dots) and GFMCSR(full dots)
calculation of $p_d$ at $J=0.4t$. (see text for details)}\label{fig:parbig}
\end{figure}
As can be seen from the comparison with the exact
results in  Fig.~\ref{fig:par18}(a), at $J=0.4t$, 
the VMC highly overestimates the order parameter. The FN reduces this value. 
The GFMCSR, implemented by reconfiguring the unperturbed energy of the
two subspaces at $N$ and $N+2$ particles in an independent way,
is almost exact.

In order to attempt a finite size scaling for the order
parameter we compute $p_d$  for much  larger sizes (Fig. \ref{fig:parbig}).
 As can be seen in the $L=50$ lattice case both the FN and the GFMCSR reduce
 the variational value. 
 We have tested  the accuracy of the calculation and
the dependency of the results from  the chosen guiding
wavefunction, by  reducing the optimal energy variational parameter
$\Delta_{DW}=0.65$ (dots connected by full lines in Fig.~\ref{fig:parbig}~a)
to the value of $\Delta_{DW}=0.3$ (dashed lines). This implies
a sizable reduction of $p_d$ within VMC. 
The FN  evaluation of $p_d$ correctly  enhances  this 
value, getting closer to the more reliable estimate obtained with 
the optimal energy variational parameter. 
The GFMCSR method, the most accurate technique used here,  
  is, remarkably, rather insensitive to the choice of the guiding function, 
 being the difference for  the two GFMCSR results a conservative  estimate of 
 the possible error in the determination of  $p_d$. GFMCSR seems to improve by 
the same amount  the  FN estimate of $p_d$, 
both for the $18$ sites (fig. \ref{fig:par18}) and $50$ sites 
(fig. \ref{fig:parbig} a), 
and this improvement is expected to remain even for larger sizes, 
being GFMCSR, as well as FN, a size consistent approximation.

 The $98$-site calculation shows that the VMC
value of $p_d$  is enhanced both by the
FN and GFMCSR calculation and remarkably the computed value is
very close to the one obtained for the $50$ site lattice.

\begin{figure}
\centerline{\psfig{figure=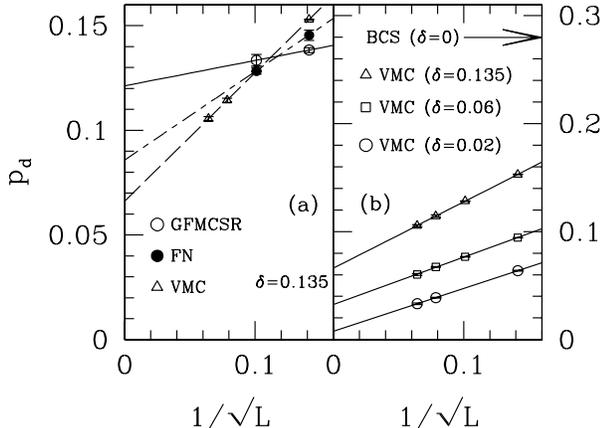,width=8cm}}
\caption{\baselineskip .185in Size scaling of $p_d$ at $J=0.4t$.
Lines connecting FN and GFMCSR in (a) are guides to the eye,
 least square fit for the   variational method in (b).}\label{fig:scaling}
\end{figure}

 Our  results   at this doping and $J/t$ value
display ,{ \em all} consistently,
   stronger and stronger  d-wave correlations, as
the accuracy of our numerical techniques are improved and 
lattice size increased.
We believe that this  represents a robust evidence
 of d-wave superconductivity in  the 2D $t-J$  model.
However the limited number of lattice sizes considered does not allow us to 
perform an accurate finite size scaling. As shown in 
Fig. \ref{fig:scaling}, size effects
are present also at the variational level and the true order parameter 
maybe below the value $\sim 0.12$ reported in the picture. 

Since the $t-J$ model originates from the doping of an antiferromagnetic Mott insulator
it is interesting to understand if the antiferromagnetic character of the undoped ground state
is resistant upon doping. Following a similar procedure to the one
 used for the superconducting long range order we added to the Hamiltonian a 
short range perturbation coupled to the 
staggered magnetization:$m_h= {1/L} \sum_R s_R^z (-)^R$ 
, namely $H \to H-h \sum_R s_R^z (-)^R$,   and compute $m_h$ in presence of the
 field $h$  either by differentiating the energy per site 
$m_h=-dE(h)/dh$ or by  using the  forward walking technique, whenever possible 
(FN).\cite{Calandra98} For this quantity the FN and GFMCSR are consistent 
for small field, meaning that the FN is already enough accurate for the 
magnetic phase diagram.

\begin{figure}
\centerline{\psfig{figure=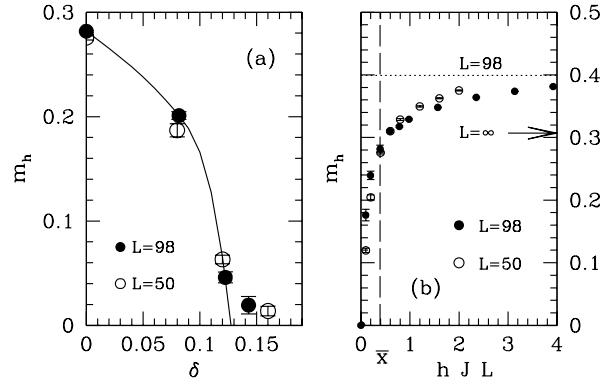,width=8cm}}
\caption{\baselineskip .185in  Staggered magnetization $m_h$ for
$\bar{ x} =\bar{h} J L=0.392$ (a).  $m_h$ for $\delta=0$ (b).
The horizontal dotted line represents the squared order parameter value.
 Remaining lines are guides to the eye.}\label{fig:afplot}
\end{figure}

\begin{figure}
\centerline{\psfig{figure=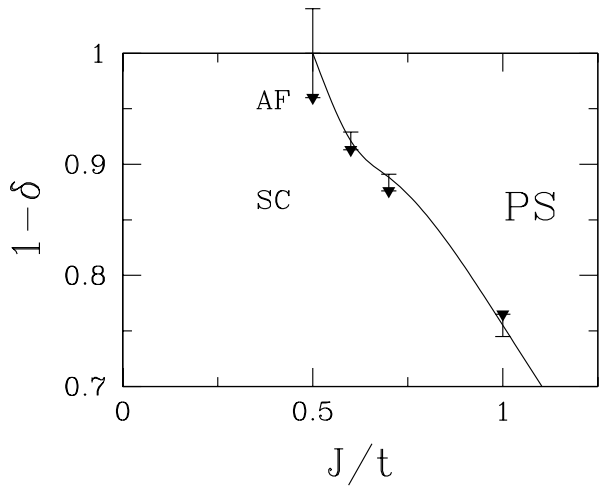,width=8cm}}
\caption{\baselineskip .185in Instability (PS) of the uniform phase 
 evaluated  by GFMCSR using the Maxwell construction for the  
$98$ site lattice. Errors are estimates of finite size effects
and correspond to twice the difference between the 98 and 50 
site critical doping \protect\cite{Calandra98b}. 
SC label the $\delta=0.14$ were $p_d$ has been computed, 
AF label the antiferromagnetic region}\label{fig:diagram}
\end{figure}

For the Heisenberg antiferromagnet, where
broken symmetry occurs, the magnetization as a function of the 
rescaled field $h \to h\cdot L$ lies on a universal curve \cite{Brezin} which 
weakly depends on the system size. This  size dependence is almost negligible
if compared to  the one  affecting  the squared order parameter 
(Fig. \ref{fig:afplot} b)\cite{Calandra98,Sandvik97}.
This feature 
strengthen the validity of our results that are {\em all}  based   
upon ground state expectation values of short range operators  in presence of a field.  
At finite doping, computationally heavier,  
we  have chosen   to work  with  a  single   field for each size and   tuned 
 at zero doping  in order   
 to reproduce on the available finite systems the infinite  size  order 
parameter: $h =  \bar x /  J L$ with $\bar x = 0.392$.
 It turns out that the
antiferromagnetic correlations are present even at finite doping up to 
$\delta_c=0.10$ see Fig.\ref{fig:afplot} (b), in qualitative
agreement with experimental findings ($\delta^{exp}_c\sim 3-5\%$). For a
quantitative agreement, other terms must be probably added to the Hamiltonian, 
as suggested in \cite{giappa}.
In the optimal doping region the staggered magnetization
 is vanishingly small 
even in presence of a sizable  magnetic field, meaning that long range order
 has disappeared favoring   a  pure $d-wave$   superconducting state. 
The interplay between antiferromagnetism and superconductivity 
appears to be a fundamental point in the phase diagram of the $t-J$ model. 
 For small doping  the  matrix element $<N+2|\Delta^+|N>$,
is strongly suppressed but antiferromagnetism  is still present. 
Close to the Mott insulator ($\delta=0$), as pointed out previously
\cite{Emery90,castellani,Hellberg97,Calandra98b},
there is a strong tendency to have a phase separation instability between an 
hole rich uniform phase and  an undoped  antiferromagnetic insulator.
In the phase diagram  shown in  Fig. (\ref{fig:diagram}), 
that we have obtained  
 with the same method (GFMCSR using only  the energy 
 in the reconfiguration scheme) used for the
 computation of the d-wave order parameter,  
 the PS boundary is quite far from the optimal doping region
 at $J/t=0.4$.
However the compressibility of the electron system is very large
 ($\frac{d \mu}{ d n} \approx 0.54 t$) 
{\em  about 20 times larger than the corresponding spinless fermion
compressibility}, in surprising numerical agreement with a spinless fermion 
model with renormalized flat band.\cite{dagotto} 
Thus the proximity to an antiferromagnetic insulator strongly enhances charge 
fluctuations determining -for physical $J/t$ values- a d-wave 
superconducting phase before the PS instability.

We believe that  a  large value of the  compressibility is very 
important to stabilize superconductivity even in presence of 
 long range Coulomb  repulsion, certainly present in the physical 
system but missing in the $t-J$  model.
For large compressibility  the Thomas-Fermi screening length
 $\xi =\frac{1}{2\pi e^2}  {d \mu  \over d n}$\cite{MarchBook} 
   is very short compared to the Cu-Cu distance, 
so that  the screening is very much effective.  We have verified 
this picture (on smaller sizes) by adding to the $t-J$ 
model a repulsive nearest-neighbor interaction $V\sum_{<ij>}n_i n_j$ 
and found still strong superconducting correlations, weakly suppressed
 even for large $V/J \sim 1$.    
 
Another mechanism in competition with   superconductivity,  
  is the formation of so called ``stripes'' in the ground state 
of the $t-J$ model,  
  as  recently   found  by White and  Scalapino with  
DMRG.\cite{White}
In order to test  this possibility we have compared our results with 
the DMRG ones  on a  $12\times6$ system with $8$ holes and open boundary 
conditions at $J/t=0.4$.
In this case, DMRG is quite more accurate than our techniques in 
the energy estimate, but it is not yet clear
whether the same is true for correlation functions especially
the ones described by short range operators, where our approximations seem to be quite reliable.
Within this accuracy for correlation functions, 
{\em  we have not found any 
clear indication of ''stripes''.}
in qualitative disagreement with the DMRG results, 
and confirming our previous work\cite{Calandra98b},
obtained with periodic boundary conditions. In this case, remarkably, 
 the possibility to use translation invariance,  allows  calculations  by far 
more accurate and reliable even compared with the best DMRG results, both 
for energies and correlation functions.

In this model we  thus recover the most simple scenario 
,  appeared  in the early days of HTc superconductivity, 
namely that the  strong correlation { \em alone}  may drive the system from 
antiferromagnetism to superconductivity.

The contradictory results
present in the literature so far are, in our opinion, mainly due to the 
general  attempt of computing a long range quantity  by using  approximations 
that weakly affect energy estimates  but may lead to  sizable  
systematic errors  on   correlation functions. 
With our technique we overcome this difficulty by estimating 
 only short range operators 
expectation values with energy difference calculations. 
The short range operators expectation values   are  less  sensitive 
to   finite size effects, and contain the  useful information 
to establish absence or presence of long range order.

We finally remark that it is extremely important  to use a very accurate 
method   to rule out superconductivity at small doping for a strongly 
correlated system like the t-J model. 
Even at the variational level the superconducting order
  parameter that is very 
large before Gutzwiller projection, becomes  an extremely small quantity  
after this projection (see Fig \ref{fig:scaling} b).
This strong suppression of $d-wave$ pairing, 
 can be easily shown at the variational level
 (see Fig.\ref{fig:scaling}) and proven at $\delta=0$, 
and is a crucial property  of strongly correlated systems.
We acknowledge S. White for  sending us numerical results before publication,
M. Fabrizio,  L. Capriotti, M. Capone, F. Becca for useful discussion 
and A. Parola for careful reading of the manuscript
This work was supported by INFM (PRA HTCS), and partly 
by MURST COFIN97 and COFIN99.

\end{document}